\documentstyle[11pt,newpasp,twoside,epsf]{article}
\def\edcomment#1{\iffalse\marginpar{\raggedright\sl#1\/}\else\relax\fi}
\reversemarginpar

\begin{document}
\title{The Large Zenith Telescope Survey: a deep survey using a 6-m liquid mirror telescope}
 \author{R\'emi A. Cabanac}
\affil{E.S.O., Alonso de Cordoba 3107, casilla 19001, Vitacura, Santiago 19, 
Chile}
\author{Paul Hickson}
\affil{U.B.C., Dept. of physics and astronomy, Hennings Bld, Vancouver, V6T 1Z4, Canada}
\author{Val\'erie de Lapparent}
\affil{I.A.P., 98 bis Bld Arago, 75014 Paris, France}
\begin{abstract}
The Large Zenith Telescope Survey whose construction is almost 
completed (first light expected in spring 2002) near Vancouver (Canada) is 
designed to observed a total strip of $\sim17\arcmin\times 120^\circ$ in 40 
narrow-band filters spanning 4000-10000 \AA. It will gather the 
spectrophotometric energy distributions of ca. $10^6$ galaxies to redshifts 
$z\sim1$, with redshift accuracy $\sigma_z=0.01$ at s/n=10, $\sigma_z=0.04$ at 
s/n=3, ca. $10^5$ stars, and a large sample of QSOs, variable stars,
and transient objects of the solar system. The survey is optimized for 
studying of the evolution of both the luminosity function and the clustering 
of galaxies to a redshift $z\sim1$. It will also provide a complete 
and homogeneous sample of stars at various galactic latitudes useful for studying  
galactic structure, and it will be a good instrument for the monitoring 
of variable objects.
\end{abstract}

Table 1. The Large Zenith Telescope in numbers\\
\begin{tabular}{llll}
\\
Latitude&49$^\circ$17\arcmin17.2\arcsec&Altitude&395 m\\
Median Seeing&0.9\arcsec&Mirror Diameter&6 m\\
CCD&Thinned 2k$\times$2k&Image scale&0.495\arcsec/pix\\
CCD width&16.9\arcmin&Area covered/night&16.9\arcmin$\times$80$^\circ$\\
Limiting R mag& 25.4&Lim. mag in filter 750 nm&24.4\\

\end{tabular}

\parbox{13cm}{\plotfiddle{filtreTC.ps}{3cm}{-90}{60}{20}{-230}{120}}
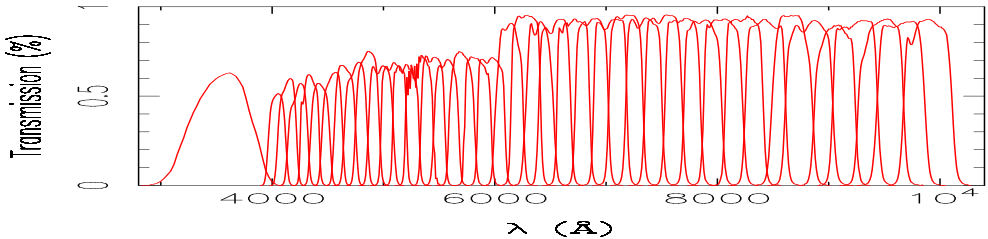
Figure 1. The transmission curves of the 40 narrow-band+U filters 
of the LZT.\\

\parbox{13cm}{\plotfiddle{lzt.eps}{8.5cm}{0}{50}{50}{-180}{-30}\\}
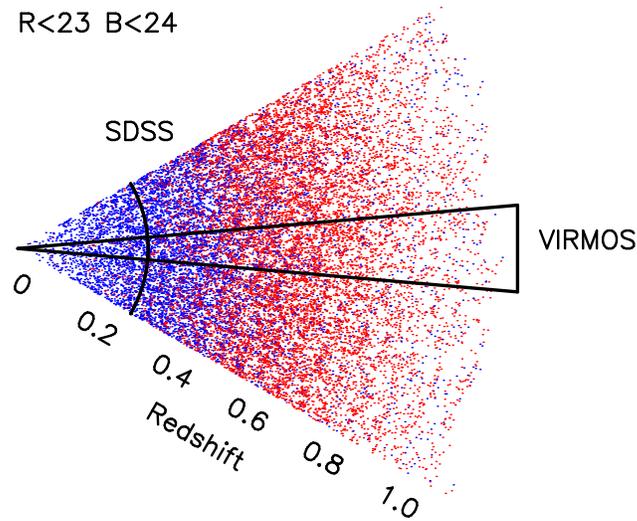
\parbox{13cm}{Figure 2. Mock LZT survey (courtesy of S. Hatton). Blue dots 
are galaxies bluer than $V-I<1.4$, red dots are galaxies redder than 
$V-I>1.4$. Overlaid are sketches of the areas of the spectroscopic surveys of SDSS and 
VIRMOS.}\\[12pt]
\parbox{5cm}{\plotfiddle{seyfert_revid.ps}{6cm}{0}{30}{30}{-95}{-25}}\hfill
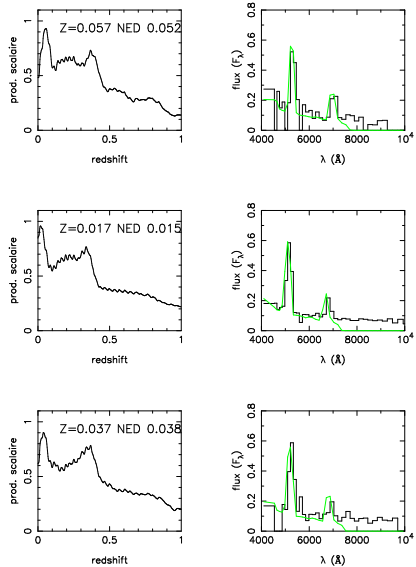
\parbox{8cm}{Figure 2. Three spectral energy distributions of HII/Seyfert 
galaxies from the UNMS1 (Hicskon \& Mulrooney 1998) obtained with the 
NASA Orbital Debris Observatory 3-m liquid mirror telescope demonstrate the 
power of narrow-band filtering. The right panels show the UNMS1 spectra
seen through 34 narrow-band filters (black 
line), with H$_\beta$+[OIII] and H$_\alpha$ emission lines clearly visible. 
The 3 objects were identified in NASA/IPAC Extragalactic Database (NED),
with names HS1314+3320, WAS69, and CASG466 (from top to bottom).
The left panels show the result of a cross-correlation with MRK 59 HII galaxy 
(green line; Kennicutt, 1992), and indicate the resulting redshifts, as well as 
the spectroscopic redshifts from NED.}

\end{document}